\documentclass[twocolumn,aps,prb,superscriptaddress]{revtex4-2}
\usepackage[T1]{fontenc}
\usepackage{babel}
\usepackage{amsmath}
\usepackage{amsfonts}
\usepackage{graphicx}
\usepackage{braket}
\usepackage{multirow}
\usepackage[unicode=true]
 {hyperref}

\makeatletter
\renewcommand{\fnum@figure}{FIG.~\thefigure}
\renewcommand{\theequation}{\arabic{equation}}
\usepackage{algorithm,algpseudocode}
\usepackage{color,xcolor}
\usepackage[caption=false]{subfig}
\definecolor{darkRed}{RGB}{130,0,0}
\definecolor{darkGreen}{RGB}{0,130,0}
\definecolor{darkBlue}{RGB}{0,0,130}
\usepackage{hyperref}
\hypersetup{
      colorlinks=true,
      citecolor=darkGreen,
      linkcolor=darkGreen,
      urlcolor=darkGreen}

\makeatother

\begin{document}

\title{Entanglement transitions in translation-invariant tensor networks}

\author{Yi-Cheng Wang}\affiliation{Department of Physics, University of California, Berkeley, CA 94720, USA}
\author{Samuel J. Garratt}
\affiliation{Department of Physics, University of California, Berkeley, CA 94720, USA}
\affiliation{Department of Physics, Princeton University, Princeton, NJ 08544, USA}
\author{Ehud Altman}\affiliation{Department of Physics, University of California, Berkeley, CA 94720, USA}
\affiliation{Materials Sciences Division, Lawrence Berkeley National Laboratory, Berkeley, CA 94720, USA}

\date{\today}

\begin{abstract}
We study the complexity of approximately contracting translation-invariant tensor networks. The computational cost of row-by-row tensor network contraction, which defines a discrete time evolution governed by a fixed transfer matrix, is associated with the entanglement of the state of a row. By analyzing a family of tensor networks whose transfer matrices interpolate between chaotic Floquet and strongly non-unitary limits, we uncover a transition between volume- and area-law entanglement in states evolved under the transfer matrix. We show that deep in the volume-law phase the spectrum of the transfer matrix in the complex plane consists of a dense ring with a sharp outer edge, reminiscent of behavior identified for non-unitary random matrices. At late times an evolving row state therefore has significant contributions from many eigenvectors with nearly degenerate eigenvalue magnitudes. In the area-law phase, there is instead a distinct leading eigenvalue. Our results establish connections between contraction complexity, spectral properties of the transfer matrix, and purification under non-unitary dynamics.
\end{abstract}

\maketitle

\emph{Introduction---}Tensor network contraction is central in computational many-body physics, underlying the representation of quantum wavefunctions~\cite{White1992density, Fannes1992finitely, Verstraete2008matrix, Schollwock2011,Cirac2021}, the simulation of quantum dynamics~\cite{Vidal2003efficient, Markov2008simulating, Kim2023evidence, Tindall2024efficient}, and the decoding of quantum error correcting codes~\cite{Ferris2014tensor, Bravyi2014efficient}. While exact contraction in two spatial dimensions is hard in the worst case~\cite{Schuch2007}, in practical row-by-row contraction one approximates the boundary state of a row by a matrix-product state (MPS) of limited bond dimension. Such an approach is efficient only when the boundary state has area-law entanglement, while it fails when the boundary state is volume-law entangled.

Remarkably, tuning bulk properties of the tensor network can lead to sharp transitions in the entanglement of these boundary states. Such phenomena have been studied extensively for random tensor networks \cite{Vasseur2019}, such as those arising in the contexts of monitored quantum dynamics \cite{Skinner2019,Li2018} and of sampling from the outputs of shallow unitary circuits \cite{Napp2022,McGinley2025}. In these random systems analytical descriptions of entanglement transitions are possible via mappings to problems in classical statistical mechanics \cite{Vasseur2019,Jian2020,Bao2020,McGinley2025}. In non-random tensor networks, on the other hand, it remains unclear whether such transitions can even occur.   

In this work we study the complexity of contracting translation-invariant tensor networks in two spatial dimensions. The structure of the boundary state is encoded in the spectral properties of a fixed transfer matrix, which can be viewed as generating non-unitary `dynamics' of a one-dimensional system of qubits, where each row of the square tensor network is associated with a different time. In the unitary limit the eigenvalues of the transfer matrix lie on the unit circle in the complex plane, and the dynamics generates a volume-law entangled state of a row. Introducing non-unitarity causes the eigenvalues to extend into a disc, and a naive expectation is that one eigenvalue will dominate at late times, just as the ground state dominates imaginary-time evolution under a quantum Hamiltonian. Here, we instead numerically demonstrate the existence of a stable volume-law entangled phase.

We argue that this stability is a consequence of the fact that, at weak non-unitarity, the disc of eigenvalues in the complex plane has a sharp outer edge. This structure ensures that, up to late times, the time-evolved row state has comparable support on many eigenvectors whose eigenvalue magnitudes are close to that of the leading one. Such a feature was previously identified in Ginibre~\cite{Ginibre1965} and deformed Haar~\cite{Wang2025} ensembles of random matrices. Increasing the degree of non-unitarity eventually `fractures' this chaotic continuum, opening a spectral gap and reducing the entanglement of the leading (left and right) eigenstates to area law. In this area-law phase we show that the row state is well-approximated by a MPS, and also that the behavior of the spectral gap can be captured using a simple product-state approximation to the leading eigenstates of the transfer matrix. We additionally show that the entanglement transition coincides with a transition in the time that it takes an initially mixed state of a row to purify \cite{Choi2020,Gullans2020dynamical}.

The tensor networks that we study in this work should be distinguished from those relevant to calculations of norms, expectation values and correlation functions in area-law entangled states of two-dimensional many-body quantum systems \cite{Schollwock2011,Cirac2021}. Such quantities involve both the quantum state and its conjugate, and as a result the corresponding row-to-row transfer matrices are $\rm{PT}$-symmetric. This additional structure drastically changes the transfer-matrix spectrum, as well as the complexity of contracting the two-dimensional network. For example, Ref.~\cite{Gonzalez2024} has recently shown that the approximate contraction of random tensor networks with this structure is computationally tractable, c.f. the volume-law entangled boundary states uncovered in Ref.~\cite{Vasseur2019}, where there is no $\rm{PT}$ symmetry. Separately, we note that Refs.~\cite{Chen2025sign,jiang2025positive} have recently shown that introducing positive bias to the tensors in a two-dimensional network (with no symmetry) drastically reduces the complexity of contraction. 

\begin{figure}[t]
\centering{}
\includegraphics[width=0.48\textwidth]{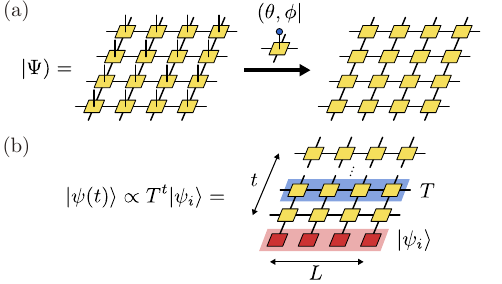}
\caption{\label{fig:KIMTN}
(a) Schematic of the scalar amplitude $(\theta,\phi|\Psi)$. The physical indices of the projected entangled pair state $|\Psi)$ are contracted with a product state $(\theta,\phi|$~(blue), reducing the system to an effective 2D tensor network with only virtual bonds. (b) Interpretation of the contracted tensor network as a time evolution. The row-to-row transfer matrix $T$ (blue shading) propagates the initial boundary state $|\psi_i\rangle$ (red shading) along the vertical temporal direction $t$, generating the effective boundary state $|\psi(t)\rangle$.
}
\end{figure}

\emph{Setup}---The contraction problem that we study arises when converting a tensor-network description of a many-qubit state into quantum state amplitudes~[Fig.~\ref{fig:KIMTN}(a)]. Consider the cluster state \cite{Raussendorf2001} on an $L \times t$ square lattice, with qubits on vertices, and with periodic boundary conditions around the $L$ direction (we consider various boundary conditions for the $t$ direction),
\begin{align}
|\Psi)=e^{-i\frac{\pi}{4}\sum_{\langle i,j\rangle}Z_iZ_j}|+)^{\otimes Lt},
\end{align}
where $|0)$ and $|1)$ are the $\pm 1$ eigenstates of Pauli $Z$ and $|+)$ is the $+1$ eigenstate of Pauli $X$. We use the notation $|\cdots)$ for states of physical degrees of freedom, and $|\cdots\rangle$ for states of the bond (or virtual) degrees of freedom which will be our focus. Although $|\Psi)$ is fully specified by a tensor with a bond dimension of just two, overlaps between $|\Psi)$ and product states encode arbitrary post-selected quantum computations~\cite{Raussendorf2001, Gross2007measurement}. Evaluating these overlaps is expected to be intractable for all physical computers \cite{aaronson2005quantum}, so it is natural to ask how and why practical methods of computation break down.

Arguably the simplest version of this problem arises when computing the overlap between $|\Psi)$ and a translation-invariant product state  $|\theta,\phi)=[\cos(\theta/2)|0)+e^{i\phi}\sin(\theta/2)|1)]^{\otimes Lt}$. The overlap of these $Lt$-qubit states can be expressed as~\cite{SM}
\begin{align}
( \theta,\phi|\Psi)&\propto\sum_{\{s\}}\text{exp}\Big[-iJ \sum_{\langle i,j\rangle}s_is_j-ih \sum_js_j\Big] \label{eq:complexsum}\\
&\propto \langle \psi_f |T^t|\psi_i\rangle. \notag
\end{align}
The first line involves a `partition function' for spins $s_i = \pm 1$ at the vertices of a square lattice, with imaginary nearest-neighbor interactions $J=\pi/4$ and a uniform complex field $h = h_R + i h_I$. The components of this field are fixed by $\theta$ and $\phi$ through $h_R=-\phi/2$ and $h_I=-\frac{1}{2}\log(\tan(\theta/2))$, meaning that the poles and equator of the Bloch sphere correspond to $h_I\to\infty$ and $h_I=0$, respectively. 

The second line of Eq.~\eqref{eq:complexsum} expresses the overlap in terms of a transfer matrix $T$ that acts on $L$-qubit rows, corresponding to bond variables of the $L \times t$ tensor network. The row states $|\psi_i\rangle$ and $\langle \psi_f|$ can be represented as $2^L$-component complex vectors, and encode fixed boundary conditions in the $t$ direction. The transfer matrix is [Fig.~\ref{fig:KIMTN}(b)]
\begin{align}
T=e^{-ig\sum_{j=0}^{L-1}X_j}e^{-iJ\sum_{j=0}^{L-1}Z_jZ_{j+1}-ih\sum_{j=0}^{L-1}Z_j},\label{eq:T}
\end{align}
and the overlap in Eq.~\eqref{eq:complexsum} can thus be evaluated by simulating the dynamics of row states of $L$ qubits for $t$ time steps, where $T$ is the evolution operator for a single `time step'. 

The couplings $J = -g = \pi/4$, fixed by the cluster state, place the $h_I=0$ dynamics of this chain precisely at a dual-unitary point of the kicked Ising model~\cite{Akila2016particle, Bertini2018SSF}. For $h_I=0$ and $h_R \neq 0,\pi/4,\pi/2$ \footnote{While for $h_R=0,\pi/2$ the unitary model can be described in terms of free fermions, for $h_R = \pi/4$ the transfer matrix $T$ is a Clifford unitary.}, the transfer matrix $T$ generates non-integrable unitary unitary evolution. The entanglement is then known to grow linearly in time when starting from a product state, saturating at volume-law scaling \cite{Bertini2019entanglement,Piroli2020exact, Zhou2022maximal}. This entanglement growth leads to the breakdown of MPS based simulations of the row-state dynamics. For $h_I \neq 0$ the tensor network is still self-dual under $\pi/2$ rotations in the $L$-$t$ plane, but both $T$ and the analogous column transfer matrix are non-unitary. We are primarily interested in the effect of increasing $h_I$ with fixed $h_R$, and focus on $h_R = \pi/6$. In the following we also restrict our analysis to the zero-momentum and reflection-symmetric sector of the space on which $T$ acts, denoting this sector by $0+$~(see Supplemental Material~\cite{SM} for other momentum sectors and open boundary condition). 

\begin{figure*}[t!]
\centering{}
\includegraphics[width=\textwidth]{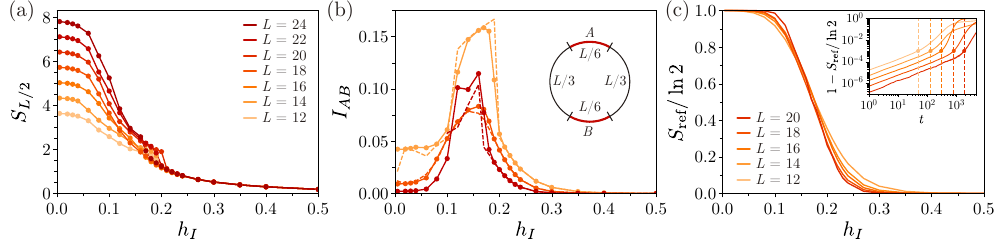}
\caption{
Entanglement and purification transitions in the non-unitary kicked Ising model, driven by increasing $h_I$ with $h_R=\pi/6$.
(a)~Half-chain entanglement entropy $S_{L/2}$ at time $t=4L$ as a function of non-unitary field strength $h_I$ for system sizes $L=12$ to $24$.
(b)~Antipodal mutual information $I_{AB}$, with region sizes $|A|=|B|=L/6$ and separation $r_{A,B}=L/2$~(inset), at time $t=4L$ as a function of $h_I$ for system sizes $L=12,18,24$. The $I_{AB}$ of the leading eigenstates, indicated by dashed lines, correspond to the infinite time limit. In (a) and (b) we average the initial row state over random translation-invariant product states. (c)~Entanglement entropy of a single reference qubit that is initially entangled with the boundary of the tensor network. The entropy is evaluated at time $t/L=0.5$ and shown as a function of $h_I$ for system sizes $L=12$ to $20$ (light to dark). A large $h_I$ drives the evolving row to an area-law entangled state and leads to rapid purification of the reference qubit, while a small $h_I$ preserves entanglement between the system and the reference qubit for a long time. Inset: Purification dynamics of the reference qubit deep within the volume-law phase, with $h_I=0.01$. The time scale $t_{\epsilon=10^{-3}}$~(dots and dashed lines), defined by $1-S_\text{ref}(t_\epsilon)/\ln2=\epsilon$, scales exponentially in system size and is associated with the earliest stages of purification. In (c) we average over pairs of random orthogonal states of the boundary (see main text).}
\label{fig:entanglement}
\end{figure*}

\begin{figure*}[t!]
\centering{}
\includegraphics[width=\textwidth]{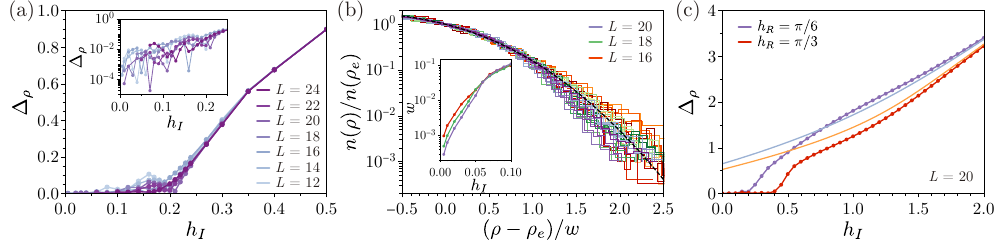}
\caption{
Spectral transition in the transfer matrix $T$. (a)~The spectral gap $\Delta_\rho\equiv\rho_0-\rho_1$ as a function of non-unitary field strength $h_I$ for even system sizes $L=12, 14, \ldots, 24$ and real field $h_R = \pi/6$. A constant gap appears at $h_I\gtrsim0.2$, while there are fluctuations and frequent crossings of leading eigenvalues at $h_I\lesssim0.2$~\cite{SM}. Inset: Gap between logarithms of the magnitudes of the leading eigenvalues at small $h_I$. For each system size $L$, the multiple kinks in $\Delta_\rho$ arise from crossings of magnitudes of the largest eigenvalues, which generically have different phases $\phi_{\alpha}$.
(b)~Collapse of radial eigenvalue density $n(\rho)$ for $L=16,18,20$ and several non-unitary fields $h_I\in[0.005,0.1]$~(increasing opacity) onto the universal complementary error function profile~(dashed line) predicted by random matrix theory [Eq.~\eqref{eq:erfc}].
Inset: The edge width $w$ extracted from the scaling collapse for system sizes $L=16,18,20$.
(c)~Comparison between numerical calculations~(dots) and mean-field ansatz~(lines) for spectral gap $\Delta_{\rho}=\rho_0-\rho_1$ at $L=20$. The strong-field regime agrees with mean-field theory, which assumes area-law entanglement and predicts an $L$-independent spectral gap.
}
\label{fig:spectral}
\end{figure*}

This work is organized as follows. In Fig.~\ref{fig:entanglement}, we will show numerically that volume-law entanglement survives over a finite window of the non-unitary perturbation $h_I\ne 0$. Beyond a critical $h_I$, the evolving state of a row undergoes an entanglement transition to an area-law phase that can be simulated efficiently using MPS. We also characterize this transition from the perspective of purification, following Refs.~\cite{Gullans2020dynamical,Choi2020}. Because the transfer matrix $T$ is fixed, this behavior must be encoded in its spectral properties, and we study how in Fig.~\ref{fig:spectral}. There we identify a transition in the gap between the magnitudes of leading eigenvalues at a critical $h_I$. We also show that, at small $h_I$ in the volume-law phase, there is a kind of random-matrix universality in eigenvalue statistics.  

\emph{Entanglement transitions---}As a first step we characterize the entanglement entropy of the time-evolved row state $T^t|\psi_i\rangle$ as well as that of the steady state of the dynamics generated by $T$, i.e. the right eigenstate whose eigenvalue has the largest magnitude.

In Fig.~\ref{fig:entanglement}(a) we compute the half-chain von Neumann entanglement entropies $S_{L/2}$ of boundary states $|\psi(t)\rangle\propto T^t|\psi_i\rangle$~ evolved for time $t=4L$, averaged over random translation-invariant product states $|\psi_i\rangle$. Throughout this work, entropies are defined using natural logarithms. As shown in Fig.~\ref{fig:entanglement}(a), $S_{L/2}$ grows linearly with system size $L$ (volume law) for weak non-unitary fields, collapsing to area-law entanglement only at $h_I \gtrsim 0.2$~\footnote{The various diagnostics provide consistent evidences of the volume-law and area-law behavior at $h_I\lesssim0.1$ and $h_I\gtrsim0.2$, respectively, while none of them gives a sharp estimate of critical point $h_{I,c}$ of the entanglement transition. Although antipodal mutual information peaks within the intermediate region $h_I\in[0.1,0.2]$, the finite-size effect and the associated crossings of leading eigenstates obscure the precise critical value.}. In the End Matter we provide evidence from infinite MPS calculations~\cite{Vidal2007classical} that, on the area-law side of this entanglement transition, the time-evolved state is well-approximated by an MPS with finite bond-dimension, and that the evaluation of the overlap in Eq.~\eqref{eq:complexsum} is computationally tractable.

To precisely locate the entanglement transition, in Fig.~\ref{fig:entanglement}(b) we study the mutual information $I_{AB}=S_A+S_B-S_{A\cup B}$ between two subregions $A,B$ on opposite sides of the system~\cite{Skinner2019,Li2019}, and at time $t=4L$. The subregions $A$ and $B$ each consist of $L/6$ sites as illustrated in the inset, where $L=12,18,24$, and we average over $\ket{\psi_i}$ as in Fig.~\ref{fig:entanglement}(a). At small $h_I$ in the volume-law phase, we find that $I_{AB}$ is exponentially small in $L$, as expected \footnote{The mutual information $I_{AB}$ in the geometry of Fig.~\ref{fig:entanglement}(b) is exponentially small in $L$ for Haar-random states. Such states provide a baseline expectation for behavior in a volume-law phase.}. At large $h_I$ in the area-law phase, $I_{AB}$ vanishes due to the exponentially decaying correlations. The entanglement transition is marked by a distinct peak in $I_{AB}$, which narrows as system size $L$ increases.

At late times when $t\to\infty$, $I_{AB}$ converges to that of the leading right eigenstate, as indicated by the dashed lines. Note that there are abrupt changes in $I_{AB}$ for $t \to \infty$ when varying $h_I$ in the volume-law phase. These changes arise from crossings of the leading eigenvalues, a hallmark of this phase which we will discuss later.

The volume-law phase that we have identified is striking when contrasted with the non-unitary kicked Ising model with $h=0$ and $g=\pm\pi/4+ig_I$. That model was previously studied in Refs.~\cite{Lu2021spacetime,Su2024dynamics}, and can be described efficiently using fermionic Gaussian states. Although a volume-law phase also exists in that case, it is unstable to arbitrarily weak non-unitary perturbation $h_I$. By contrast, the volume-law entanglement observed in our model is protected by unitary scrambling and generally requires exponential resources to simulate.

\emph{Purification transitions---}In the volume-law phase at small $h_I$, quantum information propagates through the tensor network. At strong non-unitarity, as we discuss below, $\braket{\psi_f|T^t|\psi_i}$ instead factorizes into a product of two terms, one depending on $\ket{\psi_i}$ and the other depending on $\bra{\psi_f}$. This dynamical signature of the entanglement transition can be analyzed by following the \emph{purification} of mixed initial row states, in analogy with studies of monitored quantum dynamics. 

Here we encode one bit of information by maximally entangling the initial row with a reference qubit, such that their state is $\frac{1}{\sqrt{2}}\left(|\psi_1\rangle|0\rangle_\text{ref}+|\psi_2\rangle|1\rangle_\text{ref}\right)$ where $|\psi_{1,2}\rangle$ are random orthogonal row states in the $0+$ sector. At time $t$ the state of the reference qubit is then
\begin{align}
    \rho_{\text{ref}}(t) \propto \frac{1}{2}\begin{pmatrix} \braket{\psi_1|(T^t)^{\dag} T^t|\psi_1} & \braket{\psi_2|(T^t)^{\dag} T^t|\psi_1} \\ 
    \braket{\psi_1|(T^t)^{\dag} T^t|\psi_2} & \braket{\psi_2|(T^t)^{\dag} T^t|\psi_2}
    \end{pmatrix},\label{eq:ref}
\end{align}
with $\text{Tr}\rho_{\text{ref}}(t)=1$. As the row evolves under $T^t$, the two initially orthogonal $L$-qubit states $T^t \ket{\psi_1}$ and $T^t \ket{\psi_2}$ develop nonzero overlaps, and the row loses memory of its initial conditions. This phenomenon is associated with the purification of $\rho_{\text{ref}}(t)$, which we quantify using the von Neumann entropy ${S_{\text{ref}}(t)=-\text{Tr}[\rho_{\text{ref}}\ln \rho_{\text{ref}}]}$. 

In Fig.~\ref{fig:entanglement}(c) we compute the average of $S_{\text{ref}}(t)$ over choices of $\ket{\psi_{1,2}}$ at time $t=L/2$ for various (even) $L$. We find that a large non-unitary field $h_I$ causes $S_{\text{ref}}(t)$ to rapidly decay to zero. This behavior is associated with the `collapse' of the time-evolved row state into an area-law entangled state [see Fig.~\ref{fig:entanglement}(a,b)]. By contrast, for small $h_I$, the reference qubit remains close to maximally mixed, $S_{\text{ref}}(t) \approx \ln 2$, for a time that grows exponentially with $L$, as we show in the inset of Fig.~\ref{fig:entanglement}(c).

Above we have provided evidence that, on increasing $h_I$, the tensor network undergoes an entanglement transition that coincides with a purification transition. At small $h_I$ the evolving row state is volume-law entangled and retains memory of its initial conditions for large $t$. At large $h_I$, the row only develops area-law entanglement, and quickly becomes independent of its initial state $\ket{\psi_i}$. 

\emph{Spectral gap transitions---} Here we show that the entanglement transition in our translation-invariant tensor network is associated with a transition in the spectrum of the transfer matrix $T$. In terms of the spectral properties of $T$, a time-evolved row state can be expressed as
\begin{align}
|\psi(t)\rangle\propto\sum_{\alpha}e^{(\rho_\alpha+i\phi_\alpha)t}|r_\alpha\rangle\langle l_\alpha|\psi_i\rangle,\text{\ \ }\rho_0\geq\rho_1\geq\cdots,
\end{align}
where $|r_\alpha\rangle$~($|l_\alpha\rangle$) is the right~(left) eigenstate of $T$ with eigenvalue $e^{\rho_\alpha+i\phi_\alpha}$ satisfying $\langle r_\alpha|r_\alpha\rangle=1$ and $\langle l_\alpha|r_\beta\rangle=\delta_{\alpha\beta}$. All of the phenomena that we have identified above must be encoded in the eigenvalues and eigenvectors; a key quantity in the following will be the radial gap $\Delta_{\rho}=\rho_0-\rho_1$ between the leading eigenvalues.

In unitary time-periodic evolution, the eigenvalues of the transfer matrix all lie on the unit circle in the complex plane, so $\Delta_{\rho}=0$. A non-unitary perturbation generally lifts this degeneracy of the eigenvalue magnitudes and could, in principle, generate a finite $\Delta_{\rho}$. In Fig.~\ref{fig:spectral}(a) we show that this does not occur. While eigenvalues indeed depart from the unit circle in our model, for sufficiently weak $h_I$ the gap $\Delta_{\rho}$ remains small, and we discuss its behavior in detail below. A finite ($L$-independent) gap $\Delta_{\rho}$ only emerges beyond a finite $h_I \gtrsim 0.2$. 

To understand the behavior at small $h_I$, it is useful to ground our analysis in studies of non-unitary random matrix ensembles. In Ref.~\cite{Wang2025} we analytically investigated weakly non-unitary transfer matrices of the form $\zeta U$, with $\zeta=e^{h_I\sum_{j=0}^{L-1} Z_j}$ and $U$ a Haar random $2^L \times 2^L$ unitary. We showed that the eigenvalues of these matrices form a sharp-edged ring in the complex plane, such that $\Delta_{\rho}$ is exponentially small in $L$. There, the average radial eigenvalue density $n(\rho)$, normalized as $\int_{-\infty}^{\infty} n(\rho) d\rho = 2^L$, takes a universal complementary-error-function form~\cite{Byun2025Ginibre}
\begin{align}
n(\rho)=n(\rho_e)\,\text{erfc}\Big(\frac{\rho-\rho_e}{w}\Big),\label{eq:erfc}
\end{align}
where $\rho_e$ and $w$ denote the size-dependent edge location and width, respectively. For the model in Ref.~\cite{Wang2025}, the edge width scales as $w\propto e^{-b(h_I)L}$ for any finite $h_I$, mirroring the scaling in the Ginibre ensemble $w\propto 2^{-L/2}$~\cite{Byun2025Ginibre}. 

In Fig.~\ref{fig:spectral}(b) we show that edge statistics of the form in Eq.~\eqref{eq:erfc} arise also for our spatially structured transfer matrices at small $h_I$. To do this we consider an ensemble of $T$ [Eq.~\eqref{eq:T}] with fixed $h_I$ and with $h_R$ normally distributed with mean $\pi/6$ and standard deviation $\pi/60$. Shifting the numerical data by their inflection points $\rho_e$, and fitting the width parameters $w$, we find a collapse of the mean radial density $n(\rho)$ across different system sizes $L$ and non-unitary fields $h_I$ onto the universal complementary-error-function form in Eq.~\eqref{eq:erfc}. 

Crucially, the extracted widths $w$ decrease exponentially with system size $L$ at $h_I\lesssim0.05$~[Inset, Fig.~\ref{fig:spectral}(b)], demonstrating emergent random-matrix universality at weak non-unitarity in a local model. Moreover, in the Supplemental Material~\cite{SM}, we show numerically that $\rho_0-\rho_e$ \footnote{The scaling of $\rho_0-\rho_e$ with $L$ is clearer than that of $\rho_0 - \rho_1$ because the magnitudes of leading eigenvalues often cross when tensor-network parameters are varied.} decreases exponentially with $L$. Although for the system sizes that we can access we only find a clear collapse for very small values of $h_I$, we anticipate that random-matrix behavior extends up to the entanglement transition at large $L$. 

One dynamical consequence of the sharp-edged spectral ring is that, at small $h_I$, purification under $T$ takes an exponentially long time. This is because a generic initial state has support on exponentially many eigenvectors whose eigenvalues are exponentially close to $\rho_0$, preventing the row state from rapidly collapsing onto the leading eigenvector. Consequently, two initially orthogonal states $T^t|\psi_{1,2}\rangle$ remain approximately orthogonal up to time $t\sim w^{-1}$, and $S_{\text{ref}}(t)$ remains large [Fig.~\ref{fig:entanglement}(c)]. Another consequence, with important implications for the complexity of contracting the tensor network, is that weakly entangled initial row states evolve into highly entangled superpositions of many eigenvectors of $T$ [Figs.~\ref{fig:entanglement}(a,b)].

\emph{Self-consistent mean-field theory---} With increasing non-unitary field, the universal spectral edge `fractures': the single sharp-edged ring breaks into multiple concentric annuli, with a single distinct leading eigenstate. We capture this regime using a self-consistent mean-field theory. For a given site of the row, we impose that its neighboring sites are in right- and left-states $|r^m\rangle$ and $\langle l^m|$, respectively~(with $\langle l^m|r^m\rangle=1$), and thereby construct an effective single-site evolution operator $T_m$~(see End Matter)
\begin{align}
T_m=\frac{e^{-iJ}}{2}U_0\pm\frac{e^{i2J}\langle U_0\rangle U_0-2i\sin(2J)\langle U_0Z\rangle U_0Z}{2\sqrt{e^{i2J}\langle U_0\rangle^2-2i\sin(2J)\langle U_0Z\rangle^2}},\label{eq:MFtransfer}
\end{align}
which is a $2 \times 2$ non-unitary matrix and the sign $\pm$ is chosen so that the leading eigenvalue of $T_m$ is maximized. Here $U_0=e^{-igX}e^{-ihZ}$ and $\langle\mathcal{O}\rangle\equiv\langle l^m|\mathcal{O}|r^m\rangle$. The orthogonal states $|r^m\rangle$ and $\langle l^m|$ are then self-consistently obtained as the right and left leading eigenstates of $T_m$, and the mean-field prediction for the spectral gap $\Delta_{\rho}$ is then obtained by the gap of the self-consistent $T_m$.

As shown in Fig.~\ref{fig:spectral}(c), the numerical gap converges to this prediction at large $h_I$. However, mean-field theory predicts a constant gap for all $h_I$, and therefore fails to capture the transition.

\emph{Discussion---}
We have studied the complexity of contracting two-dimensional translation-invariant tensor networks, focusing on the structure of the space-evolving row state (or boundary state), and on spectral properties of the corresponding transfer matrix $T$. In particular, we have shown that the evolving row state is highly entangled over a finite window of non-unitary fields $h_I$, and that in this window the leading eigenvalues of $T$ form a dense ring in the complex plane. At small $h_I$ a weakly entangled initial row state evolves into an entangled superposition of eigenstates of $T$ whose eigenvalues have similar magnitudes. This mechanism can cause the contraction of a translation-invariant tensor network to be exponentially costly.

The spectral mechanism identified here should be distinguished from the mechanism underlying entanglement transitions in random tensor networks, which is typically understood via mappings to statistical mechanics enabled by randomness \cite{Vasseur2019,Bao2020,Jian2020}. Moreover, unlike volume-law phases identified in non-unitary circuits that are spacetime dual to unitary ones ~\cite{Ippoliti2021postselction, Lu2021spacetime, Ippoliti2022fractal}, our model exhibits a volume-law phase with no hidden unitary structure: it is self-dual under spacetime rotation, and for $h_I \neq 0$ is non-unitary in \emph{both} temporal and spatial directions. Unlike the spectral transition identified in PT-symmetric non-Hermitian dynamics in Ref.~\cite{Sarang2021}, which had a simple mean-field analog, here a related mean-field theory breaks down as the transition is approached from the area-law phase.   

While self-duality has allowed us to restrict our attention to one direction of contraction (i.e. from row to row), we do not expect that the existence of the volume-law phase depends on this symmetry. In particular, based on our results here as well as in Ref.~\cite{Wang2025}, we expect that a sharp-edged spectral ring is a defining feature of the highly entangled phase in non-unitary and translation-invariant tensor network contraction. Examining the stability of such a gapless ring in a broader class of models would deepen our understanding of the complexity of tensor-network contraction and provide connections to random matrix theory and quantum chaos. 

\emph{Acknowledgments---}The authors are grateful to Adam Nahum and Sarang Gopalakrishnan for useful discussions. This work was supported by the Gordon and Betty Moore Foundation (SJG), the NSF QLCI program through Grant No. OMA-2016245 (EA), and a Simons Investigator Award (EA).

\bibliography{ref_TITN.bib}

\section*{End Matter}

Here we investigate the entanglement transition using infinite MPS, and also by computing R\'{e}nyi entanglement entropies with index $n < 1$, which characterize the ability of MPS to accurately describe the row state. Following this we provide additional details about our self-consistent mean-field theory for the area-law phase. 

\emph{Computational tractability transition---}
In the main text we showed that the von Neumann entanglement entropy undergoes a transition from volume-law to area-law scaling as $h_I$ is increased. However, a row state with area-law scaling of the von Neumann entropy can still be computationally hard to represent if its Schmidt values exhibit a long tail. Such a tail would necessitate a large bond-dimension description. Here, we demonstrate from two complementary perspectives that these area-law entangled boundary states can be efficiently represented by MPS, i.e. with finite bond dimension.

\begin{figure}[t!]
\centering{}
\includegraphics[width=0.48\textwidth]{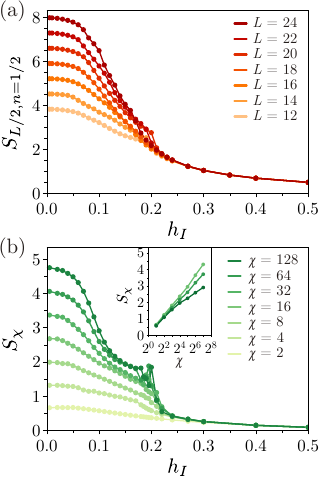}
\caption{\label{fig:KIMMPS}
Computational tractability transition in finite-size and infinite systems.
(a)~Half-chain R\'enyi-$0.5$ entropy $S_{0.5}$ of the long-time-evolved state~($t=4L$) at several system sizes $L$ as a function of the non-unitary field $h_I$.
(b)~Half-chain von Neumann entropy $S_\chi$ of the infinite chain at time $t=100$ for several bond dimensions $\chi$ as a function of $h_I$.
Inset: At $h_I=0.06, 0.08, 0.10$~(increasing opacity) in the volume-law phase, $S_\chi$ scales linearly in $\ln\chi$.
}
\end{figure}

In Fig.~\ref{fig:KIMMPS}(a) we show that, in finite-size systems, the R\'enyi entanglement entropy $S_{L/2,n}=\frac{1}{1-n}\ln \text{Tr}[\sigma_{L/2}^n]$ of the (normalized) half-chain reduced density matrix $\sigma_{L/2}=\text{Tr}_{L/2}|\psi(t)\rangle\langle\psi(t)|$ follows an area law for $h_I \gtrsim 0.2$. Here we focus on index $n=1/2$, and find an entanglement transition at approximately the same $h_I$ as for the von Neumann entropy [Fig.~\ref{fig:entanglement}(a)]. As in Fig.~\ref{fig:entanglement}(a) the state $\ket{\psi(t)} \propto T^t \ket{\psi_i}$, with $t=4L$, and we average $S_n$ over random translation-invariant product states $\ket{\psi_i}$. Our results confirm that the area-law entangled states at $h_I\gtrsim 0.2$ are dominated by a number of Schmidt values of order unity, and so can be accurately approximated by finite-bond-dimension MPS.

Second, we perform infinite MPS calculations to benchmark the volume-to-area entanglement transition in the thermodynamic limit. The transfer matrix $T$ is an MPO with bond dimension two, and the time-evolved row state $\ket{\psi(t)}$ is described by a single repeated tensor with bond dimension $\chi$. In this scheme, the entanglement scaling of time-evolved boundary state is reflected in its dependence on this virtual bond dimension $\chi$.

As shown in Fig.~\ref{fig:KIMMPS}(b), the von Neumann entanglement entropy at small $h_I$ follows an $S\sim\ln\chi$ scaling, consistent with the volume-law scaling identified in finite-size systems. By contrast, the $\chi$-independent entanglement at $h_I\gtrsim 0.2$ confirms that there is an efficient finite-bond-dimension description in the area-law phase. We note also that the area-law entanglement in this infinite-system case is exactly half of that in the finite-size system. This is because, in the infinite chain, there is a single entanglement cut, whereas our finite-size numerics have periodic boundary conditions and therefore two entanglement cuts.

\emph{Self-consistent mean-field theory---}
To construct a product state ansatz for the leading eigenstate of our transfer matrix $T$, we utilize its matrix product operator~(MPO) representation. We express the global transfer matrix as a trace over the virtual~(bond) degree of freedom, $T=\text{tr}(M^L)$, where the local tensor $M$ is
\begin{align}
M=\begin{pmatrix}
\cos(J)U_{0} & \cos(J)U_{0}Z
\\
-i\sin(J)U_{0}Z & -i\sin(J)U_{0} \label{eq:Mmatrix}
\end{pmatrix},
\end{align}
with the rows and columns of this matrix representation corresponding to the states of the bond degree of freedom, i.e. bonds connecting \emph{columns} of the two-dimensional tensor network. The entries of the matrix $M$ in Eq.~\eqref{eq:Mmatrix} are operators that evolve qubits between rows of the two-dimensional tensor network. In particular, $U_{0}=e^{-igX}e^{-ihZ}$ evolves a qubit from row to row. 

The mean-field single-site transfer matrix $T_m$ is obtained by contracting the remaining $L-1$ sites of the MPO with a product-state ansatz. Defining single-qubit states $\ket{r^m}$ and $\bra{l^m}$, with $\braket{l^m|r^m}=1$, we have
\begin{align}
T_m = \lim_{L\to\infty} \text{tr}\Big(M \Big[ \lambda_m^{-1}\langle l_m| M| r_m\rangle\Big]^{L-1}\Big), \label{eq:Tmendmatter}
\end{align}
where e.g. the upper-left component of the $2 \times 2$ matrix $\langle l_m| M| r_m\rangle$ is $\langle l_m|\cos(J)U_{0}| r_m\rangle$; in the main text we have used a compressed notation $\langle U_{0}\rangle = \langle l_m|U_{0}| r_m\rangle$ for such expectation values. The object $T_m$ can be represented as a $2 \times 2$ matrix that evolves a qubit from row to row. Taking the limit $\lim_{L\to\infty}\prod_{j=1}^{L-1} \lambda_m^{-1}\langle M \rangle$ we find the projector onto the leading eigenspace of the matrix $\langle M\rangle$. 

After constructing this projector, we can solve for $\ket{r_m}$ and $\bra{l_m}$ self-consistently. In practice, instead of finding a closed-form expression for $T_m$, we construct this object via fixed-point iteration. In particular, beginning from random guesses for the expectation values $\langle U_{0}\rangle$ and $\langle U_0Z\rangle$, we construct the associated mean-field transfer matrix, and then use its leading left and right eigenstates to update these expectation values. Iterating this procedure to convergence yields a self-consistent mean-field transfer matrix $T_m$. 

\clearpage
\pagebreak
%\onecolumngrid

\setcounter{equation}{0}
\setcounter{figure}{0}
\renewcommand{\theequation}{S\arabic{equation}}
\renewcommand{\thefigure}{S\arabic{figure}}
\renewcommand{\thesection}{SM\arabic{section}}

\begin{center}
{\large \bf Supplemental Material% for “Entanglement transitions in translation-invariant tensor networks”
}
\end{center}

\section{Amplitude of cluster state on a square lattice}\label{sec:SIamplitude}
In this section we discuss the overlap $( \theta,\phi|\Psi)$ of the cluster state $|\Psi)$ on a square $L \times t$ lattice with the translation-invariant product state $|\theta,\phi)=[\cos\frac{\theta}{2}|0)+e^{i\phi}\sin\frac{\theta}{2}|1)]^{\otimes Lt}$. The cluster state can be expressed as
\begin{align}
|\Psi)=\prod_{\langle i,j\rangle}CZ_{ij}|+)^{\otimes Lt},\label{Seq:cluster}
\end{align}
where $CZ$ represents the controlled-$Z$ gate acting on vertex qubits $j$ adjacent to $i$. Inserting resolutions of the identity $\prod_j\sum_{s_j=\pm1}|s_j)( s_j|$ into Eq.~(\ref{Seq:cluster}), the overlap $( \theta,\phi|\Psi)$ is
\begin{align}
( \theta,\phi|\Psi) &\propto\sum_{\{s\}}\prod_{\langle i,j\rangle}(1-is_is_j)\prod_j(\theta,\phi|s_j)( s_j|+)\nonumber
\\
&\propto\sum_{\{s\}}\text{exp}\Big[-iJ\sum_{\langle i,j\rangle}s_is_j-ih\sum_js_j\Big],\label{Seq:overlap}
\end{align}
where the field
\begin{align}
	h=h_R+ih_I=\frac{i}{2}\ln\Big(e^{i\phi}\cot\frac{\theta}{2}\Big).
\end{align}
We then express the overlap in terms of powers of a $2^L \times 2^L$ transfer matrix $T$ that acts on a state of $L$ virtual qubits arranged along a row, $( \theta,\phi|\Psi)=\langle\psi_f|T^t|\psi_i\rangle$, where $\langle\psi_f|$ and $|\psi_i\rangle$ represent spatial boundary conditions of the two-dimensional tensor network. The result is the Floquet evolution operator for a kicked quantum Ising model in one spatial dimension,
\begin{align}
	T=e^{-ig\sum_jX_j}e^{-iJ\sum_jZ_jZ_{j+1}-ih\sum_jZ_j},
\end{align}
where $J=-g=\pi/4$, and $j=0,1,\ldots,L-1$ labels states along a row. Note that we restrict ourselves to periodic boundary conditions in space, so we identify $Z_L \equiv Z_0$. The transfer matrix $\tilde{T}$ which acts on $t$-qubit columns has the same structure as the row-to-row transfer matrix $T$, i.e. the non-unitary transfer matrix is self-dual under spatial rotation.

\section{Translation-invariant kicked Ising model}
In this section, we discuss the conditions to observe a non-trivial entanglement transition that is not driven by symmetry. Both a nonzero Ising interaction $J\neq0$ and a real longitudinal field $h_R$ other than a measure-zero set of special points are crucial. In the absence of the Ising interaction~($J=0$), the transfer matrix decouples into a product of single-site operators, $T=U_0^{\otimes L}$ with $U_0=e^{-igX}e^{-ihZ}$, leading to a trivial area-law phase for $h_I \neq 0$, as well as a system-size-independent spectral gap. To study a generic volume-to-area-law entanglement transition, we found empirically that it is necessary to have both $J \neq 0$ and $h_R$ such that the unitary dynamics with $h_I=0$ is quantum chaotic.

\subsection{Simulable points}
The kicked Ising model $T$ possesses several integrable points in the unitary limit ($h_I=0$). The most trivial case is $h_R=0$, where the $X$ and $ZZ$ terms can be re-expressed in terms of operators quadratic in fermions using a Jordan-Wigner transformation. Other nontrivial integrable points occur at $h_R=n\pi/2$ for integer $n$, corresponding to free fermions with staggered couplings. Furthermore, for $h_R=m\pi/4$ with integer $m$, the Floquet operator is a Clifford unitary (it evolves Pauli strings into individual Pauli strings, rather than linear combinations of such strings). Although such points exhibit volume-law entanglement, they are efficiently simulable via classical algorithms. To study a nontrivial volume-to-area-law entanglement transition that reflects a transition in computational tractability, we choose $h_R$ away from these measure-zero sets. This choice guarantees that the unitary dynamics are chaotic and require exponential classical resources to simulate.

\subsection{Symmetry}
It is worth mentioning the special case  $J=\pi n/4$ with $n$ integer. For arbitrary values of the parameters (including $J$), the model has the following `symmetry'\begin{equation}
T^{-1}=VTe^{2iJ\sum_jZ_jZ_{j+1}}V^{-1},
\end{equation}
where $V=e^{-ig\sum_jX_j}\prod_jX_j\prod_jZ_j$ is a unitary operator. For $J=\pi n/4$ with $n$ integer $e^{2iJ\sum_jZ_jZ_{j+1}}=i^{nL}$ contributes a global phase. So, in this case applying the symmetry to an eigenstate $T|\lambda\rangle=\lambda|\lambda\rangle$ yields $T(V|\lambda\rangle)=i^{-nL}\lambda^{-1}V|\lambda\rangle$. This implies that the eigenvalues $\lambda_{\alpha}$ of $T$ appear in pairs satisfying $\lambda_\alpha\lambda_\beta=i^{-nL}$. Consequently, the entire eigenspectrum, in this case, must be symmetric under the inversion $|\lambda_{\alpha}|\to|\lambda_{\alpha}|^{-1}$, which translates to a $\rho_{\alpha} \to-\rho_{\alpha}$ symmetry in the radial component $\rho_{\alpha}\equiv \ln |\lambda_{\alpha}|$. Although this imposes a global constraint on the spectrum, it does not affect the local level statistics at the spectral edge. Therefore, as long as $h_R$ and $h_I$ are not set to integrable values, the edge fluctuations remain in the universality class of the complex Ginibre ensemble (class A)~\cite{Kawabata2019}.

\begin{figure}[t!]
\centering{}
\includegraphics[width=0.48\textwidth]{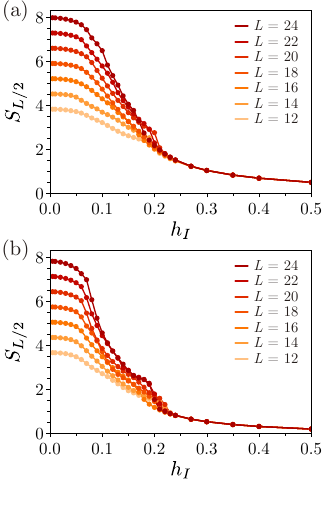}
\caption{\label{fig:Bdry_EE}
Boundary-condition dependence of the entanglement transition in the non-unitary kicked Ising model. Half-chain entanglement entropy $S_{L/2}$ at time $t=4L$ under (a) periodic and (b) open boundary conditions, for system sizes $L=12$ to $24$~(increasing opacity).
}
\end{figure}

\begin{figure}[t!]
\centering{}
\includegraphics[width=0.48\textwidth]{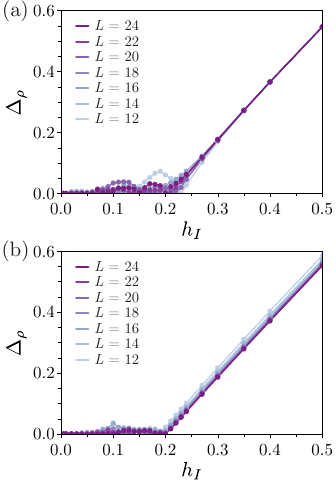}
\caption{\label{fig:Bdry_Gap}
Boundary-condition dependence of the spectral gap transition in the non-unitary kicked Ising model. Spectral gap $\Delta_\rho=\rho_0-\rho_1$ under (a) periodic and (b) open boundary conditions, for system sizes $L=12$ to $24$~(increasing opacity).
}
\end{figure}

\subsection{Boundary conditions}
In the main text, we focus on the zero-momentum, reflection-symmetric ($0+$) sector. Here we present numerical results for the von Neumann entanglement entropy $S_{L/2}$~[Fig.~\ref{fig:Bdry_EE}] and the spectral gap $\Delta_\rho$~[Fig.~\ref{fig:Bdry_Gap}] accounting for all momentum sectors. We also examine the same quantities under open boundary conditions (OBC).

To evaluate the half-chain entanglement entropy $S_{L/2}$ of the time-evolved states, we consider random initial states evolving under transfer matrices subject to both PBC and OBC. As shown in Fig.~\ref{fig:Bdry_EE}, $S_{L/2}$ exhibits the same qualitative behavior as in the $0+$ sector: the time-evolved states are volume-law entangled at small $h_I$ and collapse into area-law entangled states for $h_I\gtrsim0.2$. Similarly, the spectral gaps remain small at small $h_I$, whereas a constant spectral gap develops for $h_I\gtrsim0.2$~[Fig.~\ref{fig:Bdry_Gap}], consistent with area-law entanglement scaling. These results demonstrate that the observed entanglement and spectral gap transitions are generic features of this non-unitary model, independent of the specific momentum sector or boundary conditions.

We note that under PBC, we can track the momentum of the leading eigenstate. We find that at small $h_I$, there can be degenerate leading eigenstates with nonzero momenta $|\pm k\rangle$ due to reflection symmetry. Therefore, to show the spectral gap associated with the purification toward (degenerate) leading eigenstates, we instead consider the separation between logarithms of the amplitudes of the (degenerate) leading eigenvalues and the next-to-leading eigenvalue in Fig.~\ref{fig:Bdry_Gap}(a).

\begin{figure*}[t!]
\centering{}
\includegraphics[width=\textwidth]{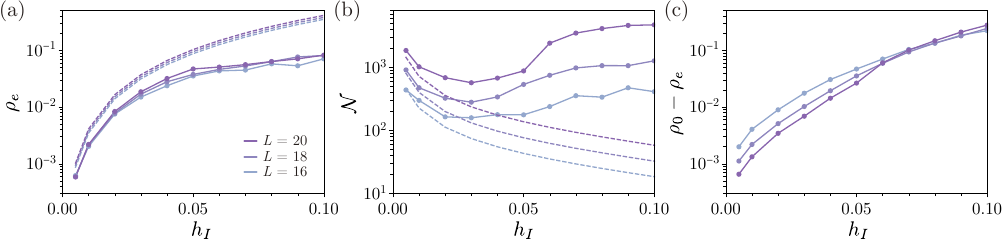}
\caption{\label{fig:KIMEdge}
Extended data for RMT universality of the radial spectral edge for the non-unitary kicked Ising model in the $0+$ sector.
(a)~Extracted edge locations as a function of non-unitary field strength $h_I$ for the non-unitary kicked Ising model~(solid lines) and the deformed Haar ensemble~(dashed lines).
(b)~Number of eigenvalues beyond the bulk edge $\mathcal{N}\equiv\int_{\rho\geq\rho_e}n(\rho)d\rho$ as a function of $h_I$ for the non-unitary kicked Ising model~(solid lines) and the deformed Haar ensemble~(dashed lines).
(c)~Averaged gap between the leading eigenvalue and the edge location as a function of $h_I$.
}
\end{figure*}

\subsection{Universal edge distribution}
In this section, we provide extended data characterizing the universal scaling properties of the spectral edge. We focus on three key metrics: the edge location $\rho_e$, the number of eigenvalues beyond the edge $\mathcal{N}$, and the gap between the leading eigenvalue and edge location.

First, the edge location is determined from the inflection point of the radial mean density $n(\rho)$. We compare the non-unitary kicked Ising model (solid lines) with the prediction from the deformed Haar ensemble (dashed lines), defined via the transfer matrix $\zeta U$ with the same non-unitary deformation $\zeta=e^{h_I\sum_jZ_j}$ but with a Haar random unitary $U$. Although the functional form of the edge profile is universal, the specific edge location $\rho_e$ is model-dependent. The extracted $\rho_e$ deviates from the deformed Haar ensemble as shown in Fig.~\ref{fig:KIMEdge}(a).

A defining characteristic of the random-matrix universality of the spectral edge is that the spectral edge contains an exponentially large number of eigenvalues. In Fig.~\ref{fig:KIMEdge}(b), we calculate the integrated spectral weight beyond the edge location, $\mathcal{N} \equiv \int_{\rho \geq \rho_e} n(\rho) d\rho$. We observe that $\mathcal{N}$ scales exponentially with system size, confirming that the edge is `dense'.

We note that at larger field strengths $h_I$, the extracted $\mathcal{N}$ exhibits non-monotonic behavior. This is likely a finite-size effect. To observe universal edge scaling, the system must be large enough to host a statistically significant number of eigenvalues within the edge width ($\mathcal{N} \gg 1$). At large $h_I$, this condition requires increasingly large system sizes $L$. Consequently, the parameter regime where we currently observe exponential scaling should be interpreted as a lower bound for the extent of the gapless phase. The true random-matrix universal region might extend to higher $h_I$ in the thermodynamic limit.

Finally, we address the possibility of outlier modes, i.e. isolated leading eigenvalues that could maintain a constant spectral gap even as the bulk edge sharpens. To rule this out, we show the average gap between the leading eigenvalue $\rho_0$ and the edge location $\rho_e$ in Fig.~\ref{fig:KIMEdge}(c). This gap vanishes exponentially with system size in the same parameter window~($h_I\lesssim0.05$) where the edge width vanishes. This confirms that the leading eigenvalue is tightly attached to the continuous spectral ring in the radial direction.

\section{Spatially-disordered kicked Ising model}
Here we study a related model to the one studied in the main text, but without translation invariance along rows. Specifically, we consider the kicked Ising model with a spatially-disordered real longitudinal field $h_{R,j}$, where the $h_{R,j}$ at different sites $j$ are drawn independently from a normal distribution with mean and standard deviation $\sigma$ both equal to $\pi/6$. In the unitary limit, this ensemble is known to be quantum chaotic for nonzero standard deviation $\sigma$ of $h_{R,j}$ \cite{Bertini2018SSF}, in the sense that its spectral statistics are described by random matrix theory. If the correspondence with random matrix theory survives with a weak non-unitary field $h_I$, then after averaging over realizations of $h_{R,j}$ we expect both a sharp-edged spectral ring and an exponentially small gap between the magnitudes of the leading two eigenvalues (as in, for example, the deformed Haar ensemble \cite{Wang2025}). We demonstrate both of these features below.

\begin{figure*}[t!]
\centering{}
\includegraphics[width=\textwidth]{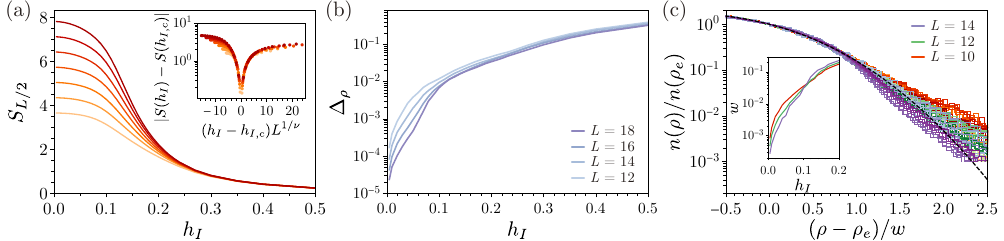}
\caption{\label{fig:KIMDis}
Entanglement and spectral properties of the non-unitary, spatially-disordered kicked Ising model.
(a)~Disorder-averaged half-chain entanglement entropy $S_{L/2}$ at time $t=4L$ as a function of non-unitary field strength $h_I$ for system sizes $L=12$ to $24$~(increasing opacity).
Inset: Scaling collapse across the critical point $h_{I,c}$, separating the volume- and area-law entanglement with $(h_{I,c},\nu)\approx(0.17,0.7)$.
(b)~Disorder-averaged spectral gap $\Delta_\rho=\rho_0-\rho_1$ as a function of non-unitary field strength $h_I$ for even system sizes $L=12$ to $18$~(increasing opacity).
(c)~Collapse of the radial eigenvalue density $n(\rho)$ for system sizes $L=10,12,14$ and several non-unitary fields $h_I\in[0.005,0.2]$ onto the universal complementary error function profile~(dashed line) predicted by random matrix theory.
Inset: The edge width $w$ extracted from the scaling collapse for system sizes $L=10,12,14$.
}
\end{figure*}

In Fig.~\ref{fig:KIMDis}(a) we numerically compute the disorder-averaged half-chain von Neumann entanglement entropy $S_{L/2}$ after evolving for $t=4L$ time steps from a random initial product state. Contrary to the translation-invariant case, disorder averaging suppresses the noisy finite-size features arising from crossings of the leading eigenvalues, thereby enabling a cleaner finite-size scaling analysis. Let us assume that the scaling form of entanglement follows
\begin{align}
\big|S_{L/2}(h_I)-S_{L/2}(h_{I,c})\big|=f\big[(h_I-h_{I,c})L^{1/\nu}\big],
\end{align}
the scaling collapse shown in the inset of Fig.~\ref{fig:KIMDis}(a) yields the critical non-unitary field $h_{I,c}\approx0.17$ with correlation length exponent $\nu\approx0.7$.

In Fig.~\ref{fig:KIMDis}(b) we show the disorder-averaged spectral gap $\Delta_{\rho}$, i.e. the separation between the logarithms of the magnitudes of the leading two eigenvalues of the transfer matrix, as a function of $h_I$ for various $L$. At small $h_I$ our results suggest $\Delta_{\rho}$ decreases exponentially on increasing $L$, as for non-unitary random matrices, while for large $h_I$ we find a very weak dependence of the gap on $L$. 

The decrease of $\Delta_{\rho}$ with $L$ at small $h_I$ suggests a sharp spectral edge, and we confirm this in Fig.~\ref{fig:KIMDis}(c). There we show that, after a rescaling of the kind discussed in connection with Fig.~\ref{fig:spectral}, the disorder-averaged radial eigenvalue density $n(\rho)$ collapses onto the universal complementary error function profile. In the inset we show that the extracted width of the spectral edge (i.e. the scale over which $n(\rho)$ decays to zero) vanishes exponentially with $L$ at $h_I\lesssim0.1$. The fact that this scaling form survives randomness in the local fields $h_{R,j}$ demonstrates that the sharp-edged spectral ring identified in the main text is not a fine-tuned feature of translation-invariant systems. Rather, it suggests that random-matrix statistics at the spectral edge are a universal feature of weakly non-unitary systems whose unitary limits are chaotic.

\end{document}